\documentclass[doublecol]{epl2} 
\usepackage[applemac]{inputenc}
\usepackage[T1]{fontenc} 
 \usepackage{amsmath} 
\usepackage{subfigure}
\usepackage{lipsum}
\usepackage{flushend}
\usepackage{amsfonts} 
\usepackage{amssymb, mathrsfs}
\usepackage{braket}
\usepackage{graphicx} 
\usepackage{color}
\usepackage{lipsum}
\usepackage{amsfonts} 
\usepackage{amssymb, mathrsfs}
\usepackage{braket}
\usepackage{bbm}
\def\beq{\begin{equation}}
\def\eeq{\end{equation}}
\def\bsp{\begin{split}}
\def\esp{\end{split}}
\def\bea{\begin{eqnarray}}
\def\eea{\end{eqnarray}}
\def\ba{\begin{array}}
\def\ea{\end{array}}

\def\lb{\left(}
\def\rb{\right)}

\def\l.{\left.}
\def\r.{\right.}

\def\ra{\rangle}
\def\la{\langle}

\def\bo{{\bf k}}

\title{Topological transitions  of spin-excitations in insulating chiral antiferromagnets}
\shorttitle{Topological transitions  of spin-excitations in insulating chiral antiferromagnets} 

\author{S. A. OWERRE\inst{1} }

\shortauthor{S. A. OWERRE}

\institute{                   
  \inst{1} Perimeter Institute for Theoretical Physics - 31 Caroline St. N., Waterloo, Ontario N2L 2Y5, Canada\\
}


\abstract{
We present a comprehensive  study of  strain-induced topological magnon phase transitions in insulating three-dimensional (3D)  topological chiral antiferromagnets on the kagome-lattice.  We show that  by applying (100) uniaxial strain,  3D insulating antiferromagnetic Weyl magnons (WMs)  manifest as an intermediate phase between a strain-induced 3D magnon Chern insulator (MCI) with integer Chern numbers  and a 3D trivial magnon insulator with zero Chern number.  In addition, we show that strain suppresses  the topological thermal Hall conductivity of magnons in these systems.  Due to the similarity between 3D insulating and metallic  kagome chiral  antiferromagnets, we envision that the current results could also manifest in the 3D antiferromagnetic topological Weyl semimetals Mn$_3$Sn\slash Ge.}

\begin{document}

\maketitle

\section{ Introduction}
Three-dimensional (3D) topological semimetals are exotic phases of matter with gapless electronic excitations, which are protected by topology and symmetry. Their theoretical predictions and experimental discoveries    have attracted considerable attention in condensed-matter physics \cite{wan,bur,xu,lv, wan1, wan2, tang, nea}. They currently remain an active field of study. Nevertheless, the   condensed matter realization of topological semimetals is essentially independent of the statistical nature of the quasiparticle excitations. In fact the notion of Weyl points was first observed experimentally in  bosonic quasiparticle excitations \cite{lu}.

\begin{figure}
\centering
\includegraphics[width=.8\linewidth]{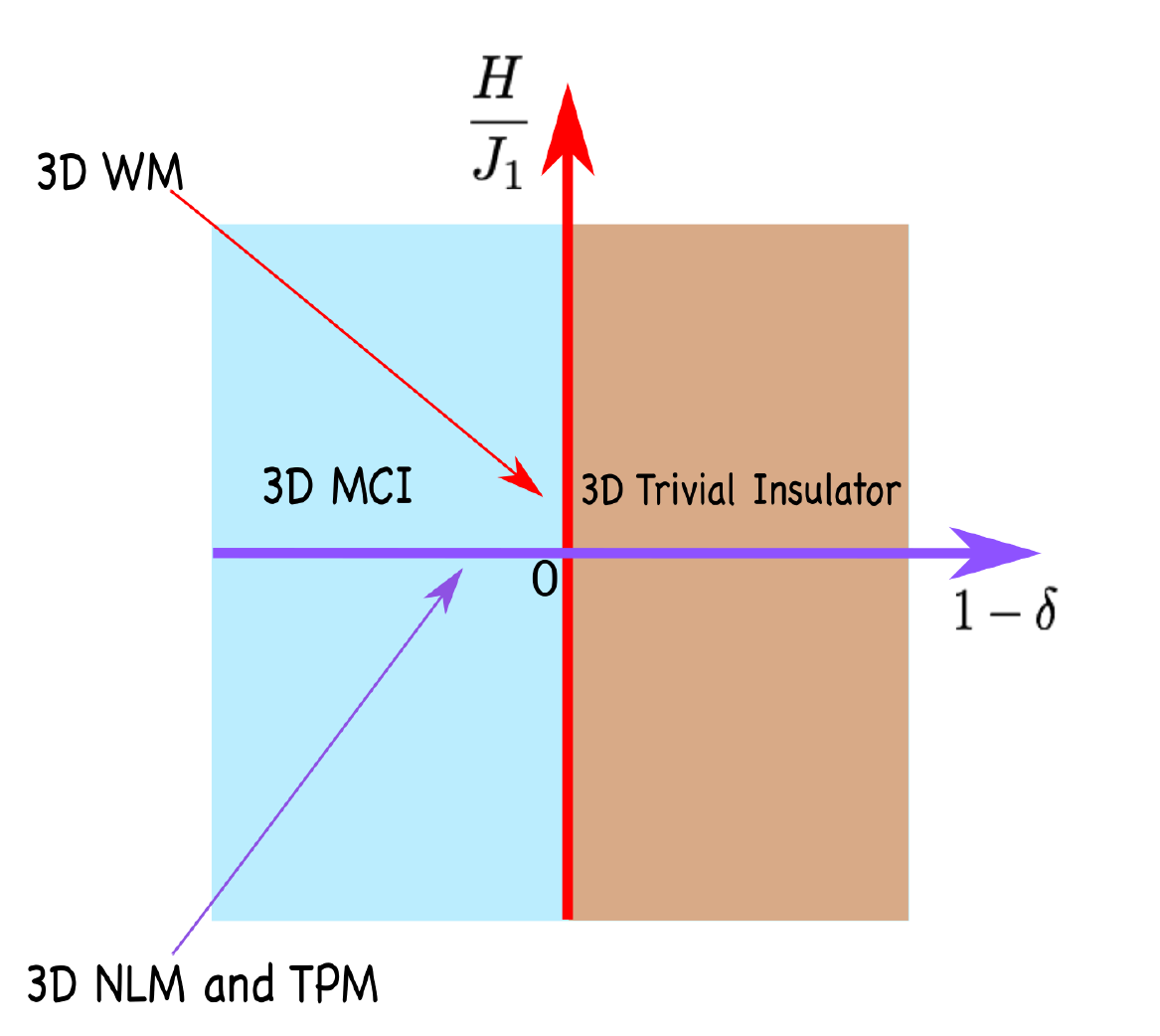}
\caption{Color online. Schematic of the topological magnon phase diagram studied in this paper. The horizontal axis represents the effect of uniaxial strain which we model as a change in the spin exchange coupling along the $(100)$ direction. The vertical axis is the magnetic field along the  $(001)$ direction. There are four magnon phases in this system. The 3D NLM and 3D TPM appear at zero magnetic field or in the conventional non-collinear spin structure. The 3D WM phase appears in the unstrained limit at nonzero magnetic field in the noncoplanar spin structure. The two new magnon  phases that appear due to strain in the noncoplanar spin structure are the 3D MCI and the 3D trivial insulator.}
\label{phase_diagram}
\end{figure}
 
 There has been an intensive search for bosonic analogs of 3D topological semimetals in insulating quantum magnets with broken time-reversal symmetry \cite{bos1, bos2, bos3, bos4, bos5, bos6, bos7, bos8, bos9, bos10, bos11, bos12, bos13}. Recently, topological Dirac magnons protected by a coexistence of inversion and time-reversal symmetry have  been experimentally observed in a 3D collinear antiferromagnet Cu$_3$TeO$_6$ \cite{bos12, bos13}.  This has opened a great avenue  for observing topological  Weyl magnon (WM) points  in 3D insulating  quantum magnets.  In magnetic bosonic systems, however, it is essentially important that the WM nodes occur at the lowest excitation if they were to make any significant contributions to observable thermal Hall transports.    This is due to the population effect of bosonic quasiparticles at low temperatures. In this respect, WM nodes at the lowest excitation can be considered as the analog of electronic Weyl points close to the Fermi energy. The  WMs in the 3D kagome chiral antiferromagnet exhibit a topological  thermal Hall effect \cite{bos7}.  Currently, they are the only  known antiferromagnetic system in which the WM nodes  occur at the lowest acoustic magnon branch and contribute significantly to the thermal Hall transports. 
 
 It is well-known that electronic ferromagnetic Weyl semimetal  occurs as an intermediate phase between an ordinary insulator and a 3D quantum anomalous Hall insulator \cite{bur,HI}. To our knowledge, this interesting topological phase boundary has not been established in 3D topological antiferromagnets. Thus far, the topological Dirac and Wely nodes that appear in 3D insulating antiferromagnets \cite{bos1, bos6, bos7, bos8, bos10, bos11, bos12, bos13} cannot transit to another topological magnon phase. In this respect, strain provides an effective way to tune the band structure of crystal in quantum materials. For instance, uniaxial strain can induce chiral anomaly and topological phase transitions in 3D topological Dirac and Wely semimetals \cite{stra, straa, strab, strac, strad, strae, straf, strag}. Moreover, strain can also induce a 3D topological Dirac semimetal in epitaxially-grown $\alpha$-Sn films on InSb(111) \cite{stra1}. We envision that such strain effects could be possible in 3D topological antiferromagnets.

In this letter,   we propose a strain-induced topological magnon phase transition  in 3D topological insulating chiral antiferromagnets.  Due to the nature of the topological magnon band distributions in the 3D kagome chiral antiferromagnet, we have chosen this system for our study. However, our results can be  extended to other 3D topological insulating antiferromagnets such as Cu$_3$TeO$_6$ \cite{bos12, bos13}. We show that under (100) uniaxial strain, a topological magnon phase transition exist in the 3D topological insulating kagome chiral antiferromagnet.  

We have identified four different magnon phases in this system as shown in Fig.~\eqref{phase_diagram}.  The 3D nodal-line magnon (NLM) and the triple point magnon (TPM) appear at zero magnetic field or at zero in-plane   Dzyaloshinskii-Moriya  interaction (DM) interaction \cite{dm,dm1} with a conventional 3D in-plane 120$^\circ$ non-collinear spin structure. They can be tuned by strain. The 3D WM phase appears in the unstrained limit at nonzero magnetic field or non-zero in-plane DM  interaction with a noncoplanar chiral spin structure. The two new magnon phases that appear due to strained noncoplanar chiral spin structure are the fully gapped 3D MCI and the fully gapped 3D trivial insulator.   We show that the former has integer Chern numbers, whereas the latter has  zero Chern number.  The study of the topological thermal Hall effect of magnons shows that the  thermal Hall conductivity is suppressed in the fully gapped insulator phases. This implies that strain suppresses the  thermal Hall conductivity of magnons.

   \begin{figure}
\centering
\includegraphics[width=1\linewidth]{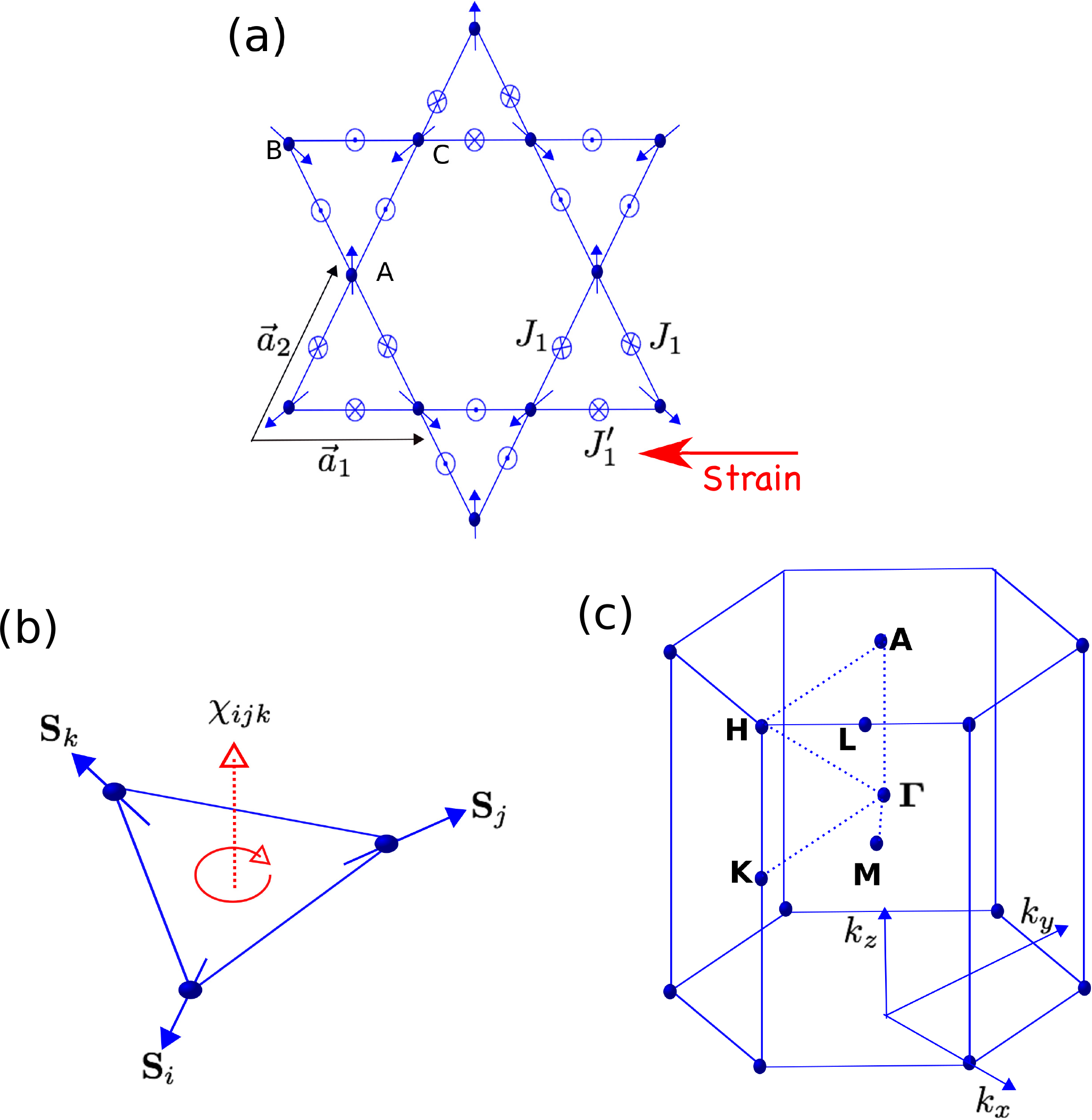}
\caption{Color online.    (a) Top view of stacked kagome-lattice  antiferromagnets  along the (001) direction. The  distribution of the DM interaction is denoted by the crossed and dotted circles. The blue arrows denote the conventional non-collinear spin order at zero magnetic field.  A uniaxial strain is applied along the $(100)$ direction. (b)  Configuration of the scalar spin chirality on the kagome lattice.   (c) The bulk Brillouin zone of the hexagonal lattices.}
\label{lattice}
\end{figure}

\begin{figure*}
\centering
\includegraphics[width=.9\linewidth]{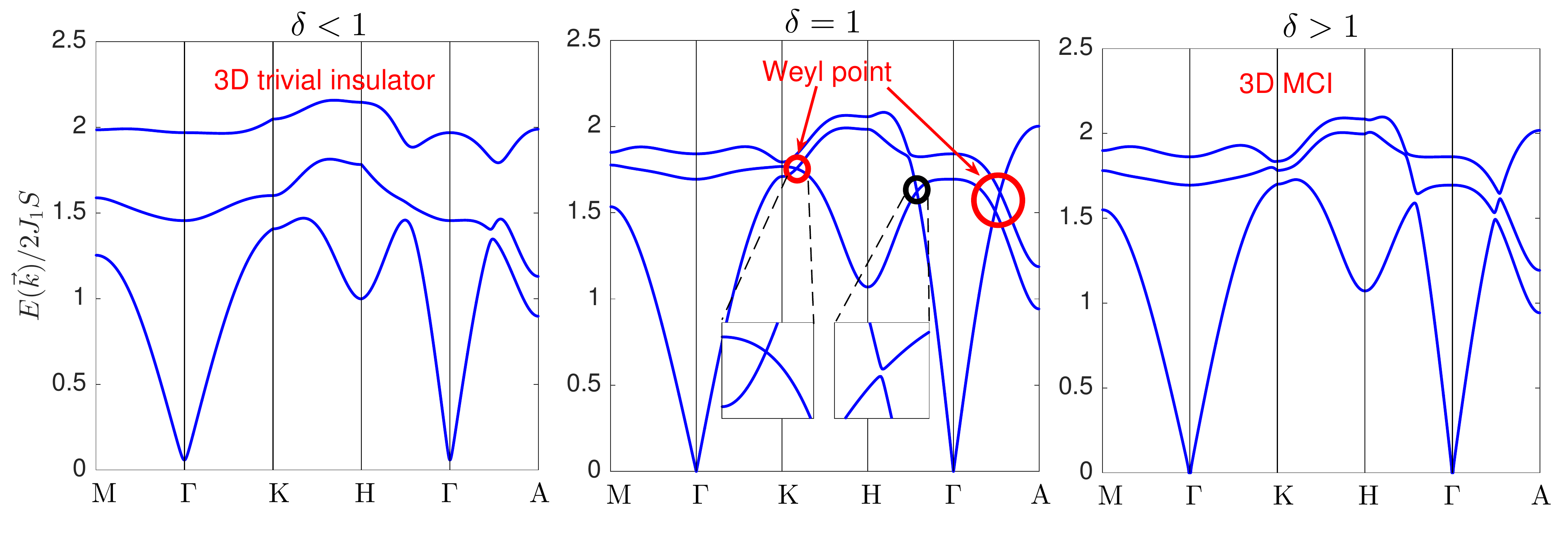}
\caption{Color online. Strain-induced topological magnon phase transitions in 3D kagome chiral antiferromagnets. (a)  3D trivial magnon insulator for $\delta = 0.75$, (b) 3D acoustic WM for $\delta = 1$, (c) 3D MCI for $\delta = 1.05$. The other parameters are set as  $J_c/J_1=0.6$, $D/J_1=0.2$, $H/J_1=0.5$.}
\label{phase_A}
\end{figure*}

\section{Strained layered  chiral antiferromagnets}
 We study  3D kagome chiral antiferromagnets in the presence of $(100)$ uniaxial strain and an external magnetic field along the $(001)$ direction. The  intralayer Heisenberg spin model in the presence of the DM interaction, interlayer coupling, and an external  magnetic field is given by
\begin{align}
\mathcal H&=\sum_{\la ij\ra,\ell}J_{ij} {\vec S}_{i}^{\ell}\cdot{\vec S}_{j}^{\ell}+\sum_{ \la ij\ra,\ell} {\vec D}_{ij}\cdot{\vec S}_{i}^{\ell}\times{\vec S}_{j}^{\ell}\nonumber\\&+J_c \sum_{\la \ell \ell^\prime\ra, i}{\vec S}_{i}^{\ell}\cdot{\vec S}_{i}^{\ell^\prime}-{\vec H}\cdot\sum_{i,\ell}{\vec S}_{i}^{\ell},
\label{model}
\end{align}
where ${\vec S}_{i}^{\ell}$ is the spin vector at site $i$ in layer ${\ell}$. The first term is the intralayer nearest-neighbour Heisenberg coupling. We model the effect of uniaxial strain  along the $(100)$ direction using the approximation that only the Heisenberg  spin interaction  along the in-plane $x$ direction changes. In this case $J_{ij}=J_1$ along the diagonal bonds and $J_{ij}=J_1\delta$ along the horizontal bonds, where $\delta$ is the strain as shown in Fig.~\ref{lattice}(a).  An alternative approximation is to consider isotropic Heisenberg interactions with lattice deformation in which only the primitive lattice vectors change. This will also modify the in-plane diagonal bonds, although to a lesser extent.  We note that since the Weyl nodes in the isotropic limit is along the out-of-plane direction, it suffices to consider only the change along the in-plane horizontal bonds as the change along the in-plane diagonal bonds will not give any new topological phase transitions. In other words, the topological phase transition requires a modification of the in-plane coupling constants.  The second term is the out-of-plane (${\vec D}_{ij}=\pm D{\hat z}$) DM interaction due to    inversion symmetry breaking between two sites on each kagome  layer. The DM interaction alternates between the triangular plaquettes of the kagome lattice and it stabilizes  the conventional in-plane $120^\circ$ non-collinear spin structure. Its sign  determines the vector chirality of the non-collinear spin order \cite{men1}.   The third term  is the nearest-neighbour interlayer antiferromagnetic coupling between the kagome  layers, which is inevitably present in real kagome materials \cite{ka,ka1,ka2,ka3,ka4}.  Finally, the last term is an external magnetic field along the stacking direction ${\vec H}= g\mu_B H{\hat z}$, where $g$ is the Land\'e g-factor and $\mu_B$ is the Bohr magneton.  

In the absence of strain, {\it i.e.} $\delta=1$, the 3D noncoplanar chiral kagome chiral antiferromagnets are intrinsic WM semimetals. The noncoplanar chiral spin texture with  macroscopically broken time-reversal symmetry can be  induced by an in-plane intrinsic DM interaction or an external magnetic field \cite{bos7}. The WM phase in this system cannot transit to any other magnon phase by changing the parameters of the system at $\delta=1$. Hence, it is strictly robust.  In this paper, we will investigate the fate of the WM phase when a uniaxial strain is applied along the $(100)$ direction. A similar effect can be induced by applying pressure. First, let us understand the spin structure at $H=0$.  In this case, the  in-plane spins on each kagome layer are canted by the angle $\varphi=\arccos(-1/2\delta)$, where $\varphi\neq 120^\circ$ for $\delta \neq 1$.  There are  various magnetic phases for different limiting cases of $\delta$. Nevertheless, our main concern here is the regime where the in-plane  spins are non-collinear and stable. This happens in the regime $\delta>1/2$.

\section{Symmetry protection of in-plane noncollinear spin structure}
\label{sym}
As previously discussed in ref.~\cite{bos7}, the conventional 3D in-plane $120^\circ$  spin structure at zero magnetic field preserves all the symmetries of the kagom\'e lattice. In particular, the combination of time-reversal symmetry (denoted by $\mathcal T$)  and spin rotation denoted by $\mathcal R_z(\pi)$ is a good symmetry \cite{suzuki}, where $\mathcal R_z(\pi)=\text{diag}(-1,-1,1)$ denotes  a $\pi$ spin rotation of the in-plane coplanar spins about the $z$-axis, and `\text{diag}' denotes diagonal elements.  In addition, the system also has three-fold rotation symmetry along the $z$ direction denoted by  $\mathcal C_{3}$. The combination  of mirror reflection symmetry $\mathcal M_{x,y}$ of the kagome plane about the $x$ or $y$ axis and $\mathcal T$ ({\it i.e.} $\mathcal T\mathcal M_x \mathcal T $ or $\mathcal M_y \mathcal T $ ) is also a symmetry of the conventional in-plane $120^\circ$ non-collinear spin structure  \cite{suzuki,ele3}. These symmetries are known as the ``effective time-reversal symmetry''  and they lead to nodal-line magnons and triply-degenerate nodal magnon points in the conventional 3D in-plane $120^\circ$  spin structure of the stacked  kagome antiferromagnets.

\section{Field-induced noncoplanar chiral spin texture}
\begin{figure*}
\centering
\includegraphics[width=1\linewidth]{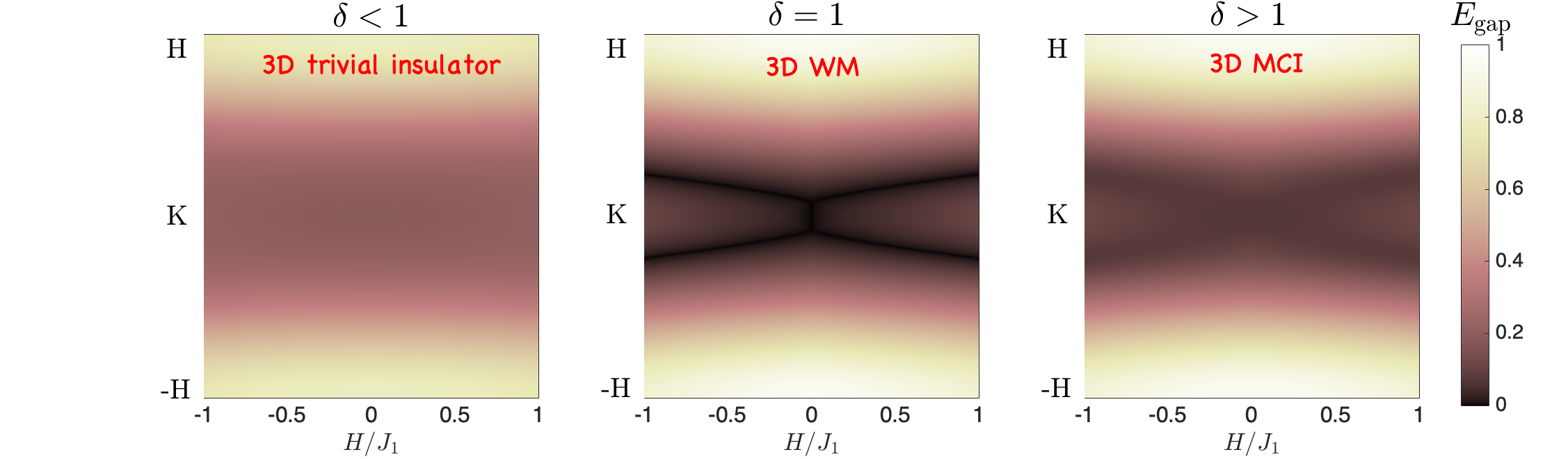}
\caption{Color online. Heat map of the energy gap between the two acoustic magnon branches as a function of the momentum along the high symmetry line -$H$--$K$--$H$ and magnetic field.  (a) 3D trivial magnon insulator for $\delta=0.75$. (b) 3D WMs for $\delta=1$.  (c) 3D MCI for $\delta =1.05$. The other parameters are set as  $J_c/J_1=0.6$, $D/J_1=0.2$.}
\label{Gap}
\end{figure*} 
 Next, we induce noncoplanar chiral spin texture in the non-collinear regime by applying a magnetic field along the $(001)$ stacking direction. We note that a  noncoplanar chiral spin texture can also be induced if the intrinsic in-plane DM interaction is present \cite{ka,ka1,ka2,ka3,ka4}. Due to the presence of an out-of-plane DM interaction, a magnetically-ordered phase is present at low temperatures. In the ordered phase the magnetic excitations are magnons (quantized spin waves). They can be captured clearly in the linear spin wave theory approximation.  First,  let us express the spins in terms of local axes, such that the $z$-axis coincides with the spin direction. This can be done by performing a local rotation $\mathcal{R}_z(\theta_{i,\ell})$ about the $z$-axis  by the spin orientated angles $\theta_{A,B,C} = 0,\varphi,-\varphi$, where $\varphi\neq 120^\circ$ for $\delta \neq 1$. As the  external magnetic field induces canting of spins in the out-of-plane direction, we perform another rotation  $\mathcal{R}_y(\chi)$ about $y$-axis  by the  angle $\chi$.  Now, the spins transform as 
 \begin{align}
{\vec S}_{i,\ell} \to\mathcal{R}_z(\theta_{i,\ell})\cdot\mathcal{R}_y(\chi)\cdot{\vec S}_{i,\ell},
\label{trans}
 \end{align}
where  the rotation matrices $\mathcal{R}_z(\theta_{i,\ell})$ and $ \mathcal{R}_y(\chi)$ are given in the Supplemental Material (SM). Using the Holstein Primakoff  transformation  \cite{hp}, the non-interacting spin-wave Hamiltonian in momentum space can be written as \begin{align}
&\mathcal H= S\sum_{\bo,\alpha,\beta}2\lb \gamma_{\alpha\beta}^{(0)}\delta_{\alpha\beta} +\gamma_{\alpha\beta}^{(1)}\rb a_{\bo \alpha}^\dagger a_{\bo \beta}\label{main}\\&\nonumber +\gamma_{\alpha\beta}^{(2)} \lb a_{\bo \alpha}^\dagger a_{-\bo \beta}^\dagger +a_{\bo \alpha} a_{-\bo \beta}\rb,
\end{align}
where $ \gamma_{\alpha\beta}^{(i)}$ are $3\times 3$ matrices (see SM),  and $S$ is the value for the spin. The three sublattices on the kagome lattice are $\alpha,\beta=A,~B,~C$.   Here, $a_{\bo\alpha}^\dagger (a_{\bo\alpha})$ are the bosonic creation (annihilation) operators.

 As shown in SM,  a finite magnetic field $H\neq 0$ induces a noncoplanar chiral spin texture with finite scalar spin chirality [see Fig.~\ref{lattice}(b)]  given by \bea 
 \chi_{ijk,\ell}=  \cos\chi{\vec S}_{i,\ell}\cdot\big( {\vec S}_{j,\ell}\times{\vec S}_{k,\ell}\big),\eea where $ \cos\chi= H/H_S(\delta)$ and the saturation field $H_S(\delta)$ is given in  SM. 
\begin{figure}
\centering
\includegraphics[width=1\linewidth]{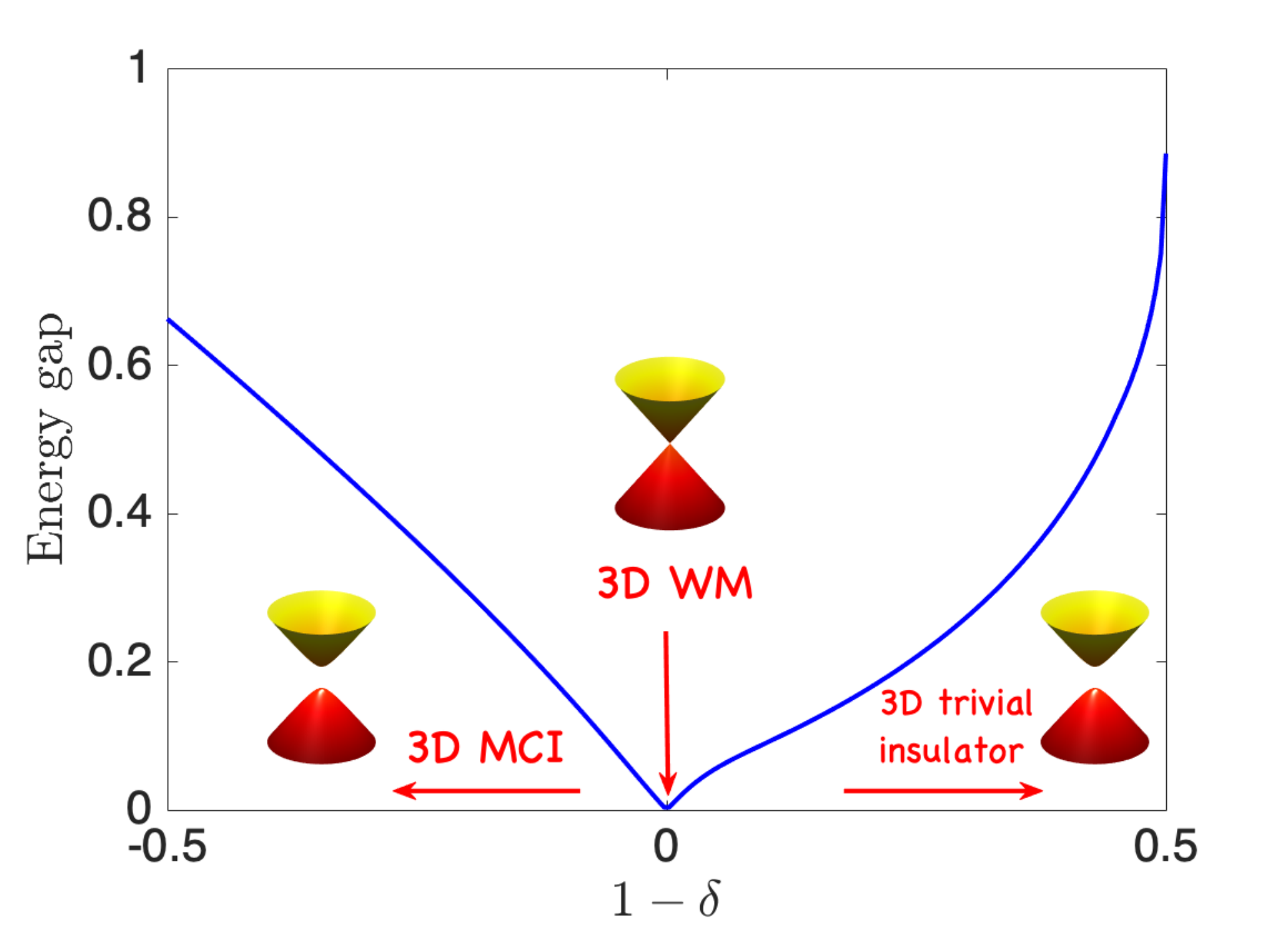}
\caption{Color online. Energy band gap between the two lowest acoustic magnon bands along $K$--$H$ high symmetry line  as a function of strain. The parameters are   $J_c/J_1=0.6$, $D/J_1=0.2$, and $H/J_1 = 0.5$.}
\label{Band_gap}
\end{figure}
In contrast to  other  magnetic Weyl systems that rely on time-reversal symmetry breaking by magnetic order, it is important to note that the scalar spin chirality breaks time-reversal symmetry macroscopically, and induces Weyl nodes at $\delta=1$.  At $H=0$ the scalar spin chirality vanishes.  In this case,  the unstrained 3D kagome chiral antiferromagnet at $ \delta =1$ exhibits a coexistence of 3D NLM and 3D TPM due to the symmetry protection of the conventional 3D in-plane $120^\circ$  spin structure at zero magnetic field as discussed above. These symmetry-protected nodal-line magnon phases can be tuned for $\delta\neq 1$ within the three magnon branches. Since an effective time-reversal symmetry is  present at $H=0$, there are no topological magnon phases in this system.  At $H\neq 0$, however, a noncoplanar chiral spin texture with macroscopically broken time-reversal symmetry is induced. As we mentioned above, the magnon bands of unstrained 3D kagome chiral antiferromagnet at $ \delta =1$ possesses robust WM points in the noncoplanar regime, and they are the only topological magnon phase in this regime.

\section{Strain-induced topological phase transitions}
 Remarkably, the strained 3D kagome chiral antiferromagnet in the noncoplanar regime exhibits a topological magnon phase transition with interesting features.  We will now investigate different aspects of these topological phase transitions. First, let us consider the 3D magnon band structures with varying $\delta$.  To establish a 3D spin structure, we fix a strong interlayer coupling $J_c/J_1=0.6$. In Fig.~\eqref{phase_A} we have shown the evolution of the  3D magnon bands along the Brillouin zone (BZ) paths in Fig.~\ref{lattice}(c).  We can see that the uniaxial strain along the $(100)$ direction gaps out the 3D WM phase at $\delta = 1$ along the high symmetry lines $K$--$H$ and $\Gamma$--$A$ of the BZ. The Goldstone mode at the $\Gamma$ point  for $\delta \geq 1$ is due to due broken continuous rotational symmetry about the $z$-axis, i.e. $U(\theta) = e^{i\theta S_z}$. The reason why there is no Goldstone mode in Fig.~\ref{phase_A}(a) is because  the chosen $\delta =0.75$ is close to the critical point $\delta =0.5$ below which there is a change of magnetic order.

\begin{figure}
\centering
\includegraphics[width=1\linewidth]{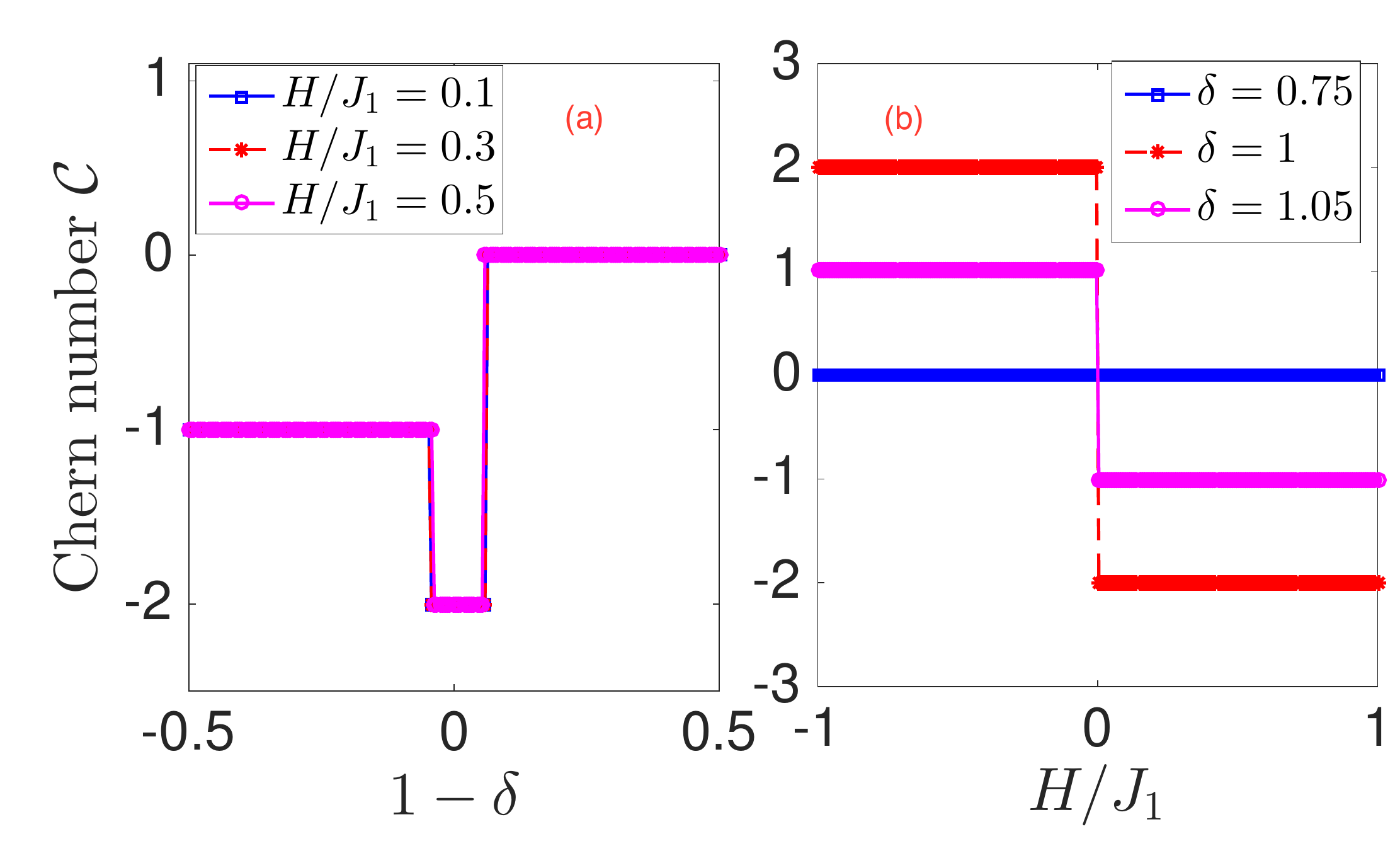}
\caption{Color online. (Left) Chern number of the lowest acoustic magnon band as a function of strain $\delta$ for different values of the magnetic field. (Right)  Chern number of the lowest acoustic magnon band as a function of the magnetic field $H/J_1$ for different values of strain $\delta$. The other parameters are set as  $J_c/J_1=0.6$, $D/J_1=0.2$.}
\label{ChernN}
\end{figure}

Let us define the gap between the two acoustic magnon branches as $E_{gap}=2|E_2(\bo)-E_1(\bo)|.$ At the WM points  $E_{gap}$ vanishes, and the fully gapped magnon insulators  are characterized by a non-zero $ E_{gap}$ along the high symmetry lines of the BZ. Next,  let us  check if the regime $\delta\neq 1$ is truly a fully gapped magnon insulator with  $ E_{gap}\neq 0$ for varying magnetic field in the noncoplanar regime. For this purpose, we  have fixed the antiferromagnetic interlayer coupling to $J_c/J_1=0.6$ and the DM interaction to $D/J_1=0.2$. We then plot the heat map of $E_{gap}$ as a function of the momentum along the high symmetry lines and the magnetic field in the noncoplanar regime. The  heat map  of  $E_{gap}$ is shown in Figs.~\ref{Gap}(a)--(c) for different regimes of $\delta$. At the critical point $\delta =1$ we can see that gapless points (black lines) appear  between the two acoustic magnon branches, which signify the presence of   WM points along -$H$--$K$--$H$ lines. In the regimes $\delta<1$ and $\delta>1$ there are no discernible gapless points along the  -$H$--$K$--$H$ line of the BZ. Therefore these two regimes define fully gapped magnon insulators, however with different properties as we will see later.  Furthermore,  in Fig.~\eqref{Band_gap} we have shown the evolution of  the magnon energy band gap $ E_{gap}$ along the $K$--$H$ high symmetry line as a function of the strain parameter $\delta$. We can see that the three distinct regions are clearly identified. We have checked that similar trends are manifested  along the -$A$--$\Gamma$--$A$ line.

Now, we will consider the Chern number topological phase transition of the system. This will justify the topological and non-topological regimes of $\delta$.  We will focus on the lowest acoustic magnon branch in which the strain-induced topological phase transition occurs.   In this case, we  can formally define the 3D trivial magnon insulator as the state where the Chern number of the lowest acoustic magnon branch vanishes, and a 3D MCI as the state with non-zero integer Chern numbers.    The 3D antiferromagnetic system can be considered as slices of 2D antiferromagnetic MCIs  \cite{th8,th9} interpolating between the $k_z=0$ and $k_z=\pi$ planes.  For an arbitrary $k_z$ point the Chern number of the magnon energy branches can be defined as
\begin{align}
\mathcal C_n = \frac{1}{2\pi}\int_{BZ} d\bo_\parallel\Omega^n_{xy}(\bo_\parallel,k_z),
\end{align}
where $\bo_\parallel= (k_x,k_y)$ is the in-plane momentum vector and $\Omega^n_{\alpha\beta}(\bo_\parallel,k_z)$ is the momentum space Berry curvature for a given magnon band $n$  as shown in  SM.  

For an arbitrary $k_z$ point and $\delta\neq 1$ the Chern number is well-defined in the noncoplanar regime.   The Chern number of all the 2D slices at an arbitrary $k_z$ point is the same because the planes at two $k_z$ points can be adiabatically connected without closing the gap for $\delta\neq 1$. For $\delta=1$ the net contribution to the Chern number comes from the regime $-k_z^c<k_z<k_z^c$, where $k_z^c$ is the location of the Weyl points along the $k_z$ momentum direction. Based on this consideration, we have shown the plots of the Chern number of the lowest acoustic magnon band as a function of strain $\delta$ for different values of $H/J_1$ in Fig.~\ref{ChernN}(a), and as a function of the magnetic field $H/J_1$ for different values of $\delta$ in Fig.~\ref{ChernN}(b). In both figures, we can see that the regime $\delta<1$ has zero Chern numbers. Therefore, the system is a 3D trivial magnon insulator for $\delta<1$. For small magnetic field in the regime  $\delta>1$, however,  the Chern number of lowest acoustic magnon band is $\mathcal C=-1$ and changes it changes sign as the sign of the magnetic field is flipped. At the 3D WM phase $\delta =1$, the Chern number is $\mathcal C=-2$ for small magnetic field.

\section{Topological thermal Hall effect}
Having identified the different magnon phases in the strained 3D kagome chiral antiferromagnet, we will now study an experimentally feasible measurement that can be performed on this system. The topological thermal Hall effect of magnons refers to the generation of a transverse thermal Hall voltage in the presence of a longitudinal temperature gradient due to the presence of noncoplanar chiral spin textures. In principle, it does not necessarily require the DM interaction provided a noncoplanar chiral spin configuration can be established, for example by adding  further nearest-neighbour interactions. Therefore, the topological thermal Hall effect of magnons is different from the conventional magnon thermal Hall effect in insulating ferromagnets which strictly requires the DM interaction \cite{th1,th2,th3,th4,th5,th6,th7, th5a}.  
  
\begin{figure*}
\centering
\includegraphics[width=0.8\linewidth]{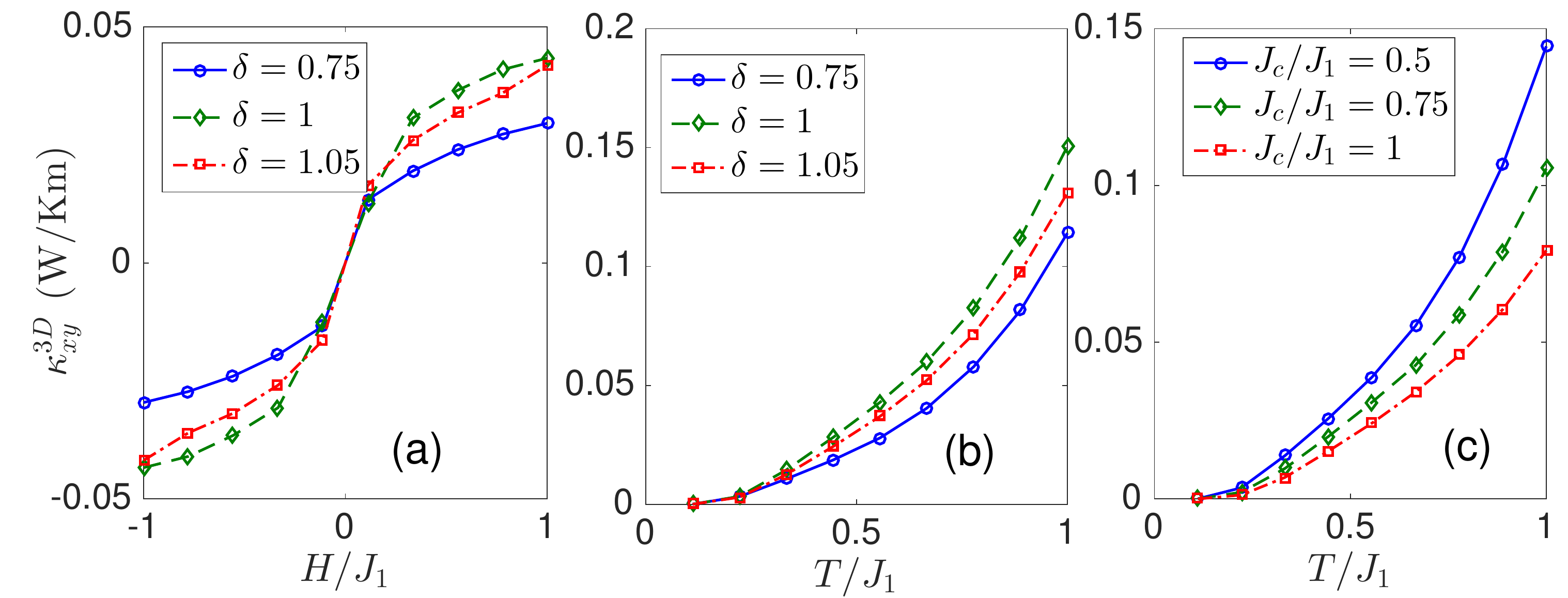}
\caption{Color online. (a) Plot of $\kappa_{xy}^{3D}$ vs.  $H/J_1$ for $T/J_1=0.5$, $J_c/J_1=0.6$, and different regimes of $\delta$.   (b) $\kappa_{xy}^{3D}$ vs.  $T/J_1$ for $H/J_1 =0.5$, $J_c/J_1=0.6$, and different regimes of $\delta$.  (c) $\kappa_{xy}^{3D}$ vs. $T/J_1$  for $H/J_1 =0.5$, $\delta=1.05$ and different values of $J_c/J_1$. The  DM interaction is   $D/J_1=0.2$ for all figures. }
\label{Hall}
\end{figure*}

\begin{figure}
\centering
\includegraphics[width=1\linewidth]{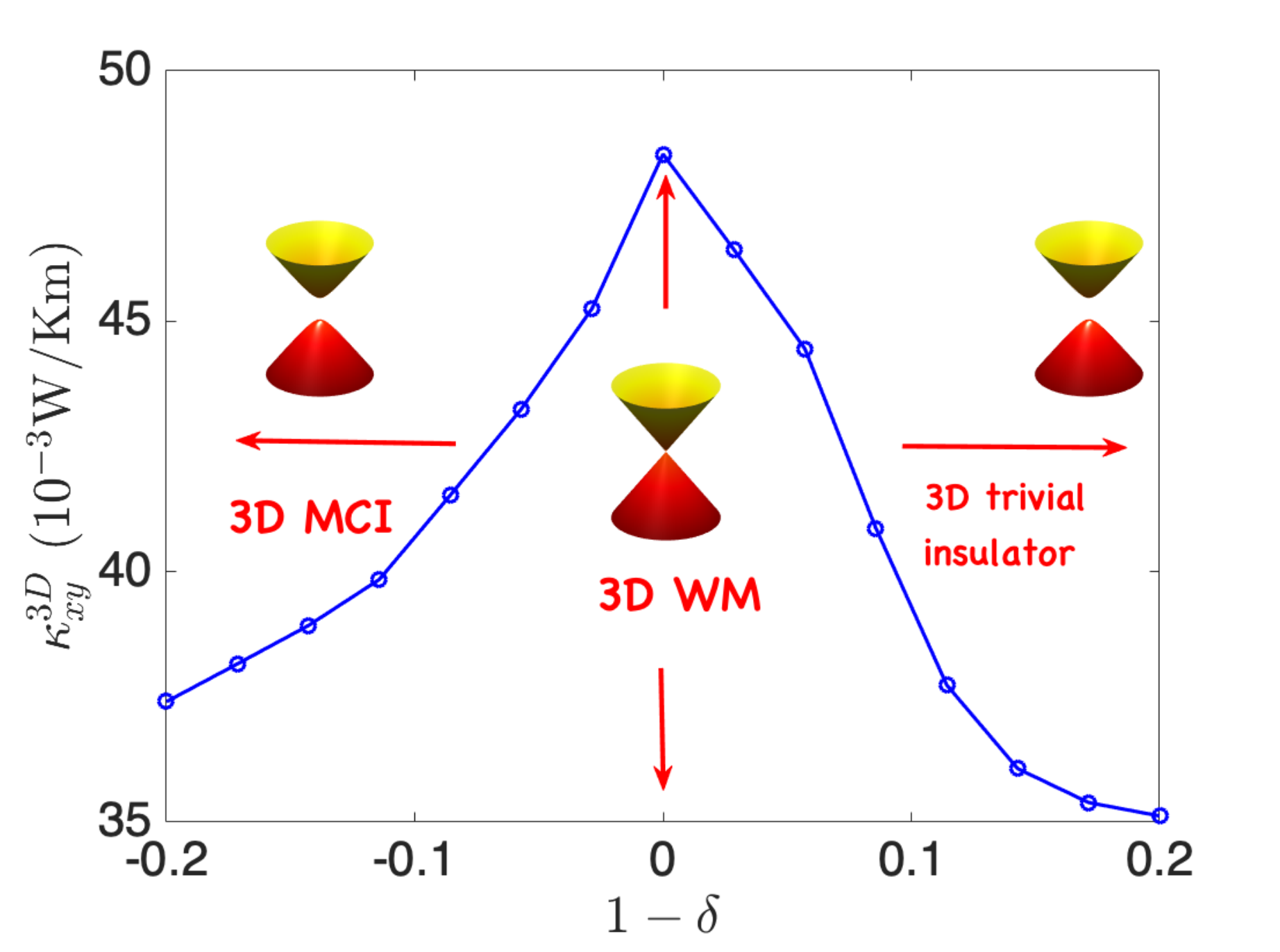}
\caption{Color online. Plot of $\kappa_{xy}^{3D}$ vs.  $1-\delta$ for $T/J_1=0.35$, $J_c/J_1=0.6$, $D/J_1=0.2$, and $H/J_1=0.5$. }
\label{THE_phase}
\end{figure}

In the 3D model, the topological thermal Hall conductivity has three contributions  $\kappa_{yz}^{3D}, \kappa_{zx}^{3D},$ and $\kappa_{xy}^{3D}$, where
the components   are  given by \cite{th5a}
\begin{align}
\kappa_{\alpha\beta}^{3D}=- k_BT\int_{BZ} \frac{d\bo}{(2\pi)^3}~ \sum_{n=1}^N c_2\lb f_n^B\rb\Omega_{\alpha\beta}^{n}(\bo),
\label{thm}
\end{align}
where   $ f_n^B=\big( e^{E_{n}(\bo)/k_BT}-1\big)^{-1}$ is the Bose occupation function, $k_B$ the Boltzmann constant which we will  set to unity, $T$ is the temperature  and $ c_2(x)=(1+x)\lb \ln \frac{1+x}{x}\rb^2-(\ln x)^2-2\text{Li}_2(-x)$, with $\text{Li}_2(x)$ being the  dilogarithm. In Eq.~\eqref{thm}, we have dropped the term $-\frac{\pi^2}{3}\sum_{n=1}^N \Omega_{\alpha\beta}^{n}(\bo)$, since the sum of the Berry curvature of the three magnon bands is zero. 

Due to the Bose occupation function, the dominant contribution to $\kappa_{\alpha\beta}^{3D}$ comes from the lowest magnon branch, where the topological phase transitions occur in the current system.  As the noncoplanar chiral spin configuration is induced along the $z$ direction, the first two components $\kappa_{yz}^{3D} $ and $ \kappa_{zx}^{3D}$ vanish. The only nonzero component is $\kappa_{xy}^{3D}$. In Figs.~\ref{Hall}(a) and (b) we have shown the trends of $\kappa_{xy}^{3D}$ as a function of the magnetic field and temperature respectively for different regimes of $\delta$. We find that the magnitude of $\kappa_{xy}^{3D}$ is suppressed in the  3D insulator phases. In other words, strain decreases the thermal Hall conductivity. We note that in the 3D WM phase the magnitude  of $\kappa_{xy}^{3D}$ is dominated by the states near the WM nodes at the lowest magnon branch due to large Berry curvatures. In this case the thermal Hall conductivity is proportional to the distance separating the WM nodes in momentum space in analogy to the anomalous Hall conductivity in Weyl semimetals \cite{bur}. In the 3D insulator phases the magnitude  of $\kappa_{xy}^{3D}$  is dominated by the states near the topological gaps at the lowest magnon branch. Despite zero   Chern number in the 3D trivial insulator phase, we can see that  $\kappa_{xy}^{3D}$ is nonzero for $\delta<1$. This is in stark contrast to electronic systems where the Fermi energy can guarantee a completely filled band and zero   Hall conductivity in the trivial insulator phase. In magnonic (bosonic) systems the Bose occupation function eliminates this restriction as the bands are thermally populated. Therefore, the thermal Hall effect in insulating quantum magnets is not a direct consequence of Chern number protected topological bands.  It depends solely on the Berry curvature of the magnon bands irrespective of their topological classifications. In Fig.~\ref{Hall}(c), we have shown the temperature dependence of $\kappa_{xy}^{3D}$ for varying interlayer coupling in the MCI phase for $\delta=1.05$. We can see that the increase in the interlayer coupling deceases $\kappa_{xy}^{3D}$.  The plot of $\kappa_{xy}^{3D}$ vs. $1-\delta$ at low temperature $T/J_1=0.35$ depicted in Fig.~\ref{THE_phase} shows the three regimes in the topological phase diagram. Evidently, the thermal Hall conductivity of Weyl magnons is maximum because they carry the dominant contribution of the Berry curvature.

\section{Conclusion}

 We have proposed a strain-induced topological phase transitions in 3D topological insulating chiral antiferromagnets.  We  showed that in the presence of (100) uniaxial strain, the  antiferromagnetic WM in 3D topological insulating antiferromagnets is an intermediate phase between a 3D antiferromagnetic MCI with integer Chern numbers and a 3D antiferromagnetic trivial insulator with zero   Chern number. We further showed that the thermal Hall conductivity of magnons is suppressed in the 3D insulator phases. Besides, we found that the 3D trivial magnon insulator with zero   Chern number possess a non-zero   thermal Hall conductivity due to the bosonic nature of magnons.  We believe that our results can be investigated experimentally in various 3D insulating antiferromagnets by applying  uniaxial strain or pressure.  For the 3D kagome chiral antiferromagnets, there are various promising materials that have been synthesized  lately \cite{ka,ka1,ka2,ka3,ka4}. Furthermore, it will be interesting to experimentally investigate the effects of strain on the recently observed 3D topological Dirac magnons in the insulating antiferromagnet Cu$_3$TeO$_6$ \cite{bos12, bos13}. As previously mentioned, we envision that the current results could also manifest in the 3D antiferromagnetic topological Weyl semimetals Mn$_3$Sn\slash Ge \cite{ele,ele1,ele2,ele3, ele4, ele5}.

\acknowledgments

Research at Perimeter Institute is supported by the Government of Canada through Industry Canada and by the Province of Ontario through the Ministry of Research
and Innovation.

\end{document}